\documentstyle[a4,11pt,graphicx]{article}

\newcommand{\mlu}{\mbox{$M_{\odot}$\,yr$^{-1}$}}
\title{
Gas and dust budget of the Large Magellanic Cloud
}
\author{
Mikako Matsuura \\
(National Astronomical Observatory of Japan; University College London)
}
\date{}

\begin{document}

\maketitle

%%% Enter your abstract here. %%%%
\begin{abstract}
 Recent  observations from the {\it Spitzer Space Telescope} enable us to study the
  mid-infrared dust excess of Asymptotic giant branch (AGB) stars in the Large Magellanic Cloud (LMC). 
  Using mid-infrared spectra, together with photometric data from the SAGE
  programme, we establish a colour selection of carbon-rich AGB stars with
  intermediate and high mass-loss rates.  We also established mass-loss rate
  versus colour relations for carbon-rich AGB stars.  The integrated
  mass-loss rate over all intermediate and high mass-loss rate carbon-rich
  AGB candidates in the LMC is $8.5\times10^{-3}$\,\mlu. This number could be
  almost doubled if oxygen-rich stars are included.
 Gas mass-loss rate from these stars is
  4--5$\times10^{-4}$\,\mlu\,kpc$^{-2}$ in the bar and
  $1\times10^{-4}$\,\mlu\,kpc$^{-2}$ outside of the bar.  
 
  AGB stars are one of the most important dust sources in the LMC, and the
  dominant gas source outside of the bar.  As a consequence of recent
  increases in the star-formation rate, supernovae are the most important
  gas source in the LMC bar and around 30\,Dor.  These differences in dust
  and gas sources impact on the gas-to-dust ratio and dust properties of the
  local ISM, because the injection from SNe could have a higher gas-to-dust
  ratio, resulting in a higher gas-to-dust ratio for the ISM in certain
  regions of the LMC.
\end{abstract}

%%% Enter your text below. %%%%
\section{Introduction}
 The interstellar medium (ISM) of a galaxy is one of the most important
factors driving its evolution.  The composition of the ISM determines the
characteristics of the next generation of stars. Simultaneously, it is
continuously being renewed and enriched by stellar ejecta. The enrichment
occurs when stars die, either exploding as supernovae (SNe) or experiencing
intense mass loss in a super-wind. Super-winds occur both in low and
intermediate mass stars during the asymptotic giant branch (AGB) phase (main
sequence masses in the range 1--8\,$M_{\odot}$), and in more massive red
supergiants. 
The origin of the gas and dust in the ISM is less well understood beyond the our Galaxy,
due to the limitation of the sensitivity.

The current rate of ISM enrichment by dust and gas depends on the total
stellar population, the initial mass function and star-forming history
(e.g. Salpeter {Salpeter55}).  Type II SNe are expected to dominate the
enrichment in the early phases of galaxy evolution, up to 100 Myrs.
It takes more than 100\,Myr
for the first intermediate-mass stars to evolve onto the AGB.  Thus, dust and gas enrichment from
AGB stars occurs later than from high-mass stars.  Different galaxies,
at different stages of this process, may be expected to show differences in
gas-to-dust ratios, dust content, and, in consequence, ISM dust extinction
curves, as well as the overall spectral distribution of galaxies.

In our Galaxy, the major dust sources are presumed to be AGB stars and SNe
(Gehrz \cite{Gehrz89}). Some other sources, such as Wolf-Rayet stars and novae, also
contribute dust to the ISM of the Milky Way, but the fraction is estimated
to be small.  The relative importance of AGB stars and SNe
remains uncertain. 
Using {\it Spitzer Space Telescope} (Werner et al. \cite{Werner04}),
now it is able to measure the gas and dust budget beyond our Galaxy.

%_________________________________________________________________
\begin{figure}[h]
\centering
\resizebox{0.5\hsize}{!}{\includegraphics*[24,289][509, 641]{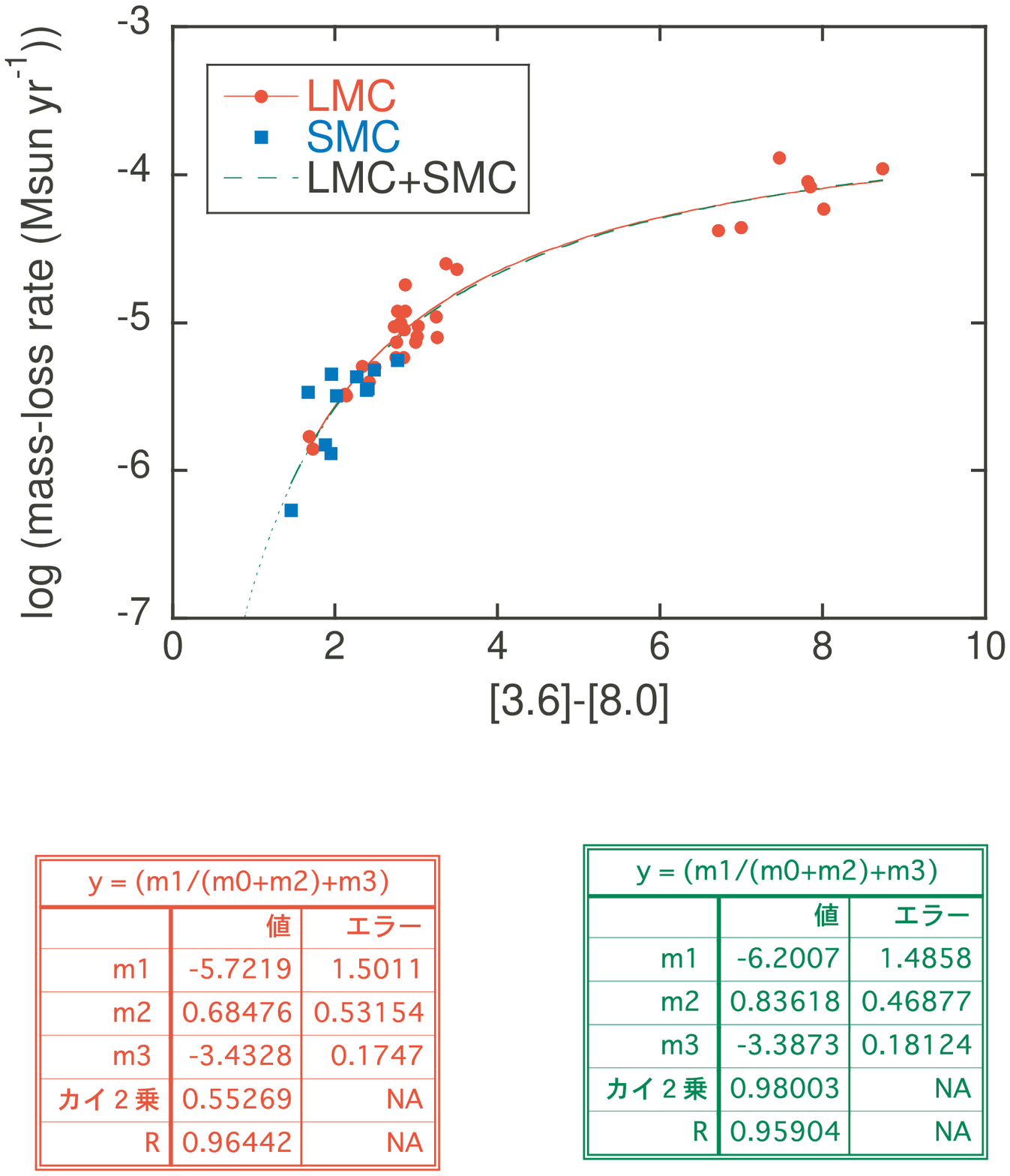}}
\caption{ Gas mass-loss rates for LMC and SMC carbon stars from
  \cite{Groenewegen07} are plotted as a function of $[3.6]-[8.0]$ colours
  extracted from SAGE (Meixner et al. \cite{Meixner06}).  The solid (red) curve is the fit to the LMC sample only and 
  the dashed (green) curve
  is the fit to the combined LMC and the SMC samples.  }
\label{Fig-3-8-massloss}
\end{figure}
%_________________________________________________________________
%_________________________________________________________________
\begin{figure}[h]
\centering
\rotatebox{90}{ 
\begin{minipage} {5.5cm}
\resizebox{\hsize}{!}{\includegraphics*[17,25][497,693]{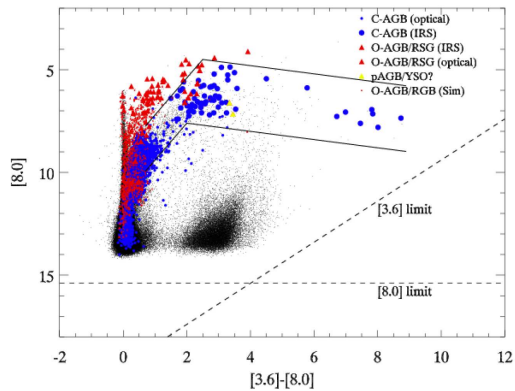}}
\end{minipage}}
\caption{
$[3.6]-[8.0]$ vs [8.0] colour-magnitude diagram.
Spectroscopically known oxygen-rich and carbon-rich AGB stars
are plotted. Selection criteria to extract carbon-rich AGB candidates
are indicated by the solid lines.
The sensitivity limits quoted by Meixner et al. \cite{Meixner06} for the final data
product are indicated by the dashed lines.
\label{Fig-38-8}}
\end{figure}
%_________________________________________________________________

\section{Analysis}

We have observed AGB stars (mainly carbon-rich stars) in the  LMC and the Small Magellanic Cloud (SMC)
(Sloan et al., Zijlstra et al. \cite{Sloan06, Zijlstra06}). All spectroscopic data are from the
Infrared Spectrograph (IRS; Houck et al. \cite{Houck04}) on-board Spitzer.  A
complete LMC photometric survey using Spitzer was recently published by
SAGE (Meixner et al. \cite{Meixner06}) and a  survey of the main part of the SMC was published by
S$^3$MC (Bolatto et al. 2007 \cite{Bolatto07}).  Combining these data, we obtain a census of the
mass-losing stars, and compare AGB stars with other dust and gas sources in
the LMC.

We derive observational mass-loss rate versus colour relations by adopting
mass-loss rates estimated for carbon-rich AGB stars in the Magellanic Clouds
from Groenewegen et al. \cite{Groenewegen07}.  
Fig.\ref{Fig-3-8-massloss} shows the mass-loss
rates as a function of [3.6]$-$[8.0].
Throughout this paper, we assume a
gas-to-dust ratio of 200 for carbon-rich AGB stars, in order to convert
the measured dust mass to a gas mass-loss rate (c.f. Matsuura et al. \cite{Matsuura07}).

To estimate the integrated mass-loss rates from carbon-rich AGB stars, it is
essential to find a classification scheme that will separate carbon- from
oxygen-rich stars (Fig.\,\ref{Fig-38-8}).  In particular, oxygen-rich AGB stars and red supergiants
(RSGs) follow different mass-loss rate vs colour relations from carbon-rich
stars and their dust contents will be very different.  

Details of these analyses are described in Matsuura et al. \cite{Matsuura08}.
%%% Enter acknowledgements here if you want. %%%
%\bigskip
%We acknowledges the anonymous referee whose helpful comments
%gave us a chance to improve the paper.
%%% Enter your reference list below. %%%

\section{Discussion}

Table\,\ref{Table-object-dist} summarizes the estimated gas and dust
production from various sources.  AGB stars are one of the main sources of
the dust enrichment for the ISM of the LMC and carbon-rich AGB stars are a
major factor.  

If a gas-to-dust ratio of 200 is valid for carbon-rich AGB stars, then SNe
make a larger contribution to the gas in the LMC bar region.  In the LMC
disk, AGB stars could be the more important source of gas and dust, due to
the low star formation activity in that region at recent times.  Other gas
and dust sources appear to present only minor contributions to the ISM
enrichment.

This is in marked contrast to our Galaxy where AGB stars are estimated to be
the main source, even for gas (Tielens et al. \cite{Tielens05}).  The difference might be
a consequence of the recent increase in the star forming rate (SFR)
 in the LMC bar region, possibly due to the tidal
interaction with the SMC.

The SFR (0.19--0.26\,$M_{\odot}$\,yr$^{-1}$) is higher than the gas injection rate from SNe and AGB stars
(in total 0.03--0.05\,$M_{\odot}$\,yr$^{-1}$).  
%estimated a SFR of 0.19\,$M_{\odot}$\,yr$^{-1}$
%, and \cite{Kennicutt95}
%suggested a SFR of 0.26\,$M_{\odot}$\,yr$^{-1}$.
%from empirical formula of H$\alpha$ luminosity of the galaxy and SFR.
These are an order of magnitude higher than the gas output from AGB
candidates and SNe.  This suggests that the LMC star formation depends on
the large reservoir of ISM gas ($7\times10^8$\,$M_{\odot}$ in H{\small I}
%(\cite[Westerlund 1997]{Westerlund97}) 
and $1\times10^8$\,$M_{\odot}$ in H$_2$). 
%(\cite[Israel 1997]{Israel97})).
Here we assume a low gas and dust inflow from the Magellanic stream. This is
probably consistent with a slow increase of the metallicity in recent times
(within the last few Gyrs), according to the age-and-metallicity relation, 
as the gas feedback from stars takes Gyr to
increase the metallicity.

%_________________________________________________________________
\begin{table*}
% \centering
  \caption{ Gas and dust injected into the ISM of the LMC}
\begin{center}
 \begin{tabular}{lrrrrrrrrccccccccc}
  \hline
Sources & Gas mass & Dust mass & Dust chemical type &  \\
 & ($10^{-6}$\,\mlu)& ($10^{-6}$\,\mlu)& \\ \hline
AGB stars \\
~~Carbon-rich                               & 8600 & 43~~ & C-rich\\
~~Oxygen-rich                               & $>>$200  & $>>$0.4 & O-rich \\
Type II SNe                                               &  20000--40000 & 0.1--130$\ddag$~ & both O- and C-rich \\
WR stars                                       & 60 & & (C-rich?)\\
Red supergiants                         & $>$1000                 & $>$2~~ & O-rich \\
\hline
\end{tabular}
\label{Table-object-dist}
\end{center}
$\ddag$ Dust production (or possibly destruction) in and around SNe remain uncertain.
% \gmas : gas mass-loss rates estimated using  $\Ks-[8.0]$ and $[3.6]-[8.0]$
\end{table*}
%________________________________________________________________


\begin{thebibliography}{}

\bibitem{Bolatto07}  
   Bolatto A.D., Simon J.D., Stanimirovi\'{c} S., et al.,
   2007, ApJ 655, 212
%\bibitem{Bowen91}
%   Bowen G.H. \& Willson L.A.,
%   1991, ApJ 375, L53
%\bibitem{Buchanan06}
%  Buchanan C.L., Kastner J.H., Forrest W.J., Hrivnak B.J.,
%  Sahai R., Egan M., Frank A., Barnbaum C.,
%  2006, AJ 132, 1890
%\bibitem{Carrera08} 
%  Carrera R., Gallart C., Hardy E., Aparicio A., \& Zinn R., 2008, AJ, 135, 836 
%\bibitem{CMR01} 
%  Chiappini C.,   Matteucci F., \& Romano D., 2001, ApJ 554, 1044 
%\bibitem{Dwek98}
%  Dwek E., 1998, ApJ 501, 643
%\bibitem{FD08} 
%  Finlator K., Dav\'e R., 2008,
%  MNRAS, 385, 2181
\bibitem{Gehrz89}
  Gehrz R., 
  1989, `Interstellar Dust', IAU Symp 135, p. 445
\bibitem{Groenewegen07}
  Groenewegen M.A.T.,  et al.,
  2007, MNRAS 376, 313
\bibitem{Houck04}
  Houck J.R., et al.,
  2004, ApJS 154, 18
%\bibitem{Israel97}
%  Israel F.P., 1997, A\&A 328, 471
\bibitem{Jura89}
 Jura M., Kleinmann S.G., 1989, 341, 359
%\bibitem{Kennicutt95}
% Kennicutt R.C, Bresolin F., B., Bomans D.J., Bothun G.D., Thompson I.B., 
% 1995, AJ 109, 594
%\bibitem{Lagadec07}
%  Lagadec E., et al.,
%  2007, MNRAS 376, 1270
\bibitem{Leisenring08}
   Leisenring J.M., Kemper F., Sloan G.C.,
  2008, ApJ 681, 1557
%\bibitem{Maeder92}
%   Maeder A., 1992, A\&A, 264, 105 
%\bibitem{Marshall04}
%  Marshall J.R., van Loon J.Th., Matsuura M., Wood P.R., Zijlstra A.A., Whitelock P.A.,
%  2004, MNRAS 355, 1348
%\bibitem{Matsuura06}
%  Matsuura M., Wood P.R., Sloan G.C., et al.,
%  2006, MNRAS 371, 415
\bibitem{Matsuura07} 
   Matsuura M., et al.,
  2007, MNRAS, 382, 1889
 \bibitem{Matsuura08} 
   Matsuura M., et al., 
   2009, MNRAS,  396, 918
%\bibitem{Mattsson08}
%  Mattsson L., Wahlin R., H\"ofner S., Eriksson K., 
%  2008, A\&A 484, L5
\bibitem{Meixner06}
 Meixner M., et al.,
  2006, AJ 132, 2268
%\bibitem{Nakamura99}
%   Nakamura T., Umeda H., Nomoto K., Thielemann F., Burrows A.,
%  1999, ApJ 517, 193
\bibitem{Salpeter55}
  Salpeter E.E., 1955, ApJ 121, 161
\bibitem{Sloan06}
   Sloan G.C., et al. 2006,
   ApJ 645, 1118
%\bibitem{Sloan08}
%   Sloan G.C., Kraemer K.E., Wood P.R., Zijlstra A.A., Bernard-Salas J., Devostand D., Houck J.R.,
%  2008,  ApJ, accepted (arXiv:0807.2998)
%\bibitem{Smecker-Hane02}  
%  Smecker-Hane T.A., Cole A.A., Gallagher J.S. III, Stetson P.B.,
%   2002, ApJ 566, 239
%\bibitem{Tielens94}
% Tielens A.G.G.M., McKee C.F., Seab C.G, Hollenbach D.J.,
% 1994, ApJ 431,  321   
\bibitem{Tielens05}
 Tielens A.G.G.M., Waters L.B.F.M., Bernatowicz T.J.,
 2005, ASPC 341, 605
%\bibitem{Vassiliadis93}
%  Vassiliadis E., \& Wood P.R., 1993, ApJ, 413, 641
%\bibitem{vanLoon00}
%  van Loon J.Th., 2000 A\&A 354, 125	
 \bibitem{Werner04}
 Werner M.W., Roellig T.L., Low F.J., et al.,
 2004, ApJS 154, 1
%\bibitem{Westerlund97}
%  Westerlund B.E., `The Magellanic Clouds', Cambridge U Press, 1997
%\bibitem{Whitney08}
%  Whitney B.A, et al.,
%  2008, AJ 136, 18
%\bibitem{Wood98}
%  Wood P.R., Habing H.J., McGregor P.J., 
%  1998, A\&A 336, 925
%\bibitem{Wood07}
% Wood P.R., Groenewegen M.A.T., Sloan G.C., Blommaert J.A.D.L.,
% Cioni M.-R.L., Feast M.W., Habing H.J., Hony S., Lagadec E.,
% 2007, ASPC 378, 251
\bibitem{Zijlstra06}
 Zijlstra A.A., et al.,
 2006, MNRAS 370, 1961
\end{thebibliography}
\end{document}